\documentclass[10pt,preprintnumbers,twocolumn,amsmath,amssymb,nofootinbib,superscriptaddress]{revtex4-1}
\usepackage[latin1]{inputenc}
\usepackage{slashed}
\usepackage{amsmath}
\usepackage{textcomp}
\usepackage{amssymb}
\usepackage{amsfonts}
\usepackage{indentfirst}
\usepackage{color}
\usepackage{hyperref}
\usepackage{subcaption}

\newcommand{\be}{\begin{equation}}
\newcommand{\ee}{\end{equation}}
\newcommand{\bea}{\begin{eqnarray}}
\newcommand{\eea}{\end{eqnarray}}

\newcommand{\nn}{\nonumber\\}

\newcommand{\mb}{\mathbf}
\newcommand{\Od}{\mathcal{O}} 

\usepackage{graphicx,graphics}
\graphicspath{{figures/}}

\begin{document}

\hfill{KCL-PH-TH/2023-01}

\title{Null energy condition violation: Tunnelling versus the Casimir effect}

\author{Jean Alexandre} 
\author{Drew Backhouse}  

\affiliation{Theoretical Particle Physics and Cosmology, King's College London, WC2R 2LS, UK}

\begin{abstract}
We show that tunnelling between two degenerate minima, as allowed in a finite volume, leads to a non-extensive symmetric ground state. This results in Null Energy Condition violation for sufficiently low temperatures, when a continuous set of momenta in the box containing the field is assumed. Taking into account discrete momenta can modify this picture and is achieved via the addition of the Casimir energy to the tunnelling-induced ground state energy. Focusing on zero-temperature, these non-trivial effects are found to compete, depending on the typical length scales involved.
\end{abstract}

\maketitle

\section{Introduction}

Spontaneous Symmetry Breaking (SSB) is strictly speaking valid for infinite volumes only, where tunnelling between degenerate vacua is completely suppressed. 
On the other hand, for a field confined in a box of finite volume, tunnelling between degenerate vacua is allowed and we study here the energetic consequences.

Involving tunnelling in the quantisation of a system automatically takes into account the different vacua and is known to lead to a convex effective action \cite{Symanzik:1969ek,Coleman:1974jh,Iliopoulos:1974ur,Haymaker:1983xk,Fujimoto:1982tc,Bender:1983nc,Hindmarsh:1985nc,Plascencia:2015pga,Millington:2019nkw}.
This is not the case in the situation of SSB, where the different vacua are decoupled and quantisation over a single vacuum does not 
necessarily lead to convexity.
Taking into account several degenerate vacua in the partition function comes with a remarkable energetic feature, generated dynamically: 
the effective action is non-extensive, as was shown in \cite{Alexandre:2012ht,Alexandre:2022sho} with a semi-classical approximation for the partition function.

The latter works were done in an $O(4)$-symmetric Euclidean spacetime though, 
and to account for a full description of tunnelling 
one needs a finite spatial volume $V$ and an independent large Euclidean time $\beta$. 
The natural context for these studies is therefore equilibrium field theory at a finite-temperature $T=1/\beta$. The corresponding Quantum Mechanics 
study was done in \cite{Alexandre:2022qxc}, involving a gas of instantons/anti-instantons which dominates the partition function in the limit of small temperature. 
It is shown there that the Null Energy Condition (NEC - see \cite{Rubakov:2014jja,Kontou:2020bta} for reviews) is violated, 
as a consequence of a non-extensive effective action induced by tunnelling. 
The present article extends this study to full 4-dimensional quantum fluctuations, and
we find that NEC violation occurs in any finite volume for sufficiently low temperatures.

Our study does not however deal with high-temperature symmetry restoration, as seen in the Kibble-Zureck mechanism \cite{Kibble:1976sj,Zurek:1985qw}.
We are instead interested in the low-temperature regime, where tunnelling dominates over thermal fluctuations providing an opportunity to violate the NEC, which the Kibble-Zurek mechanism does not.

We first evaluate quantum corrections with continuous momentum for fluctuations above each saddle point, to describe the fundamental dynamical mechanism induced by tunnelling. We then take into account the modification arising from
discrete momentum in a finite volume, using results known from studies of the Casimir effect (see \cite{Bordag:2001qi} for a review). 
The latter is known to be either attractive or repulsive, depending on the geometry of the box containing the field, as well as the 
boundary conditions the field satisfies on the walls of the box. 
As a consequence, as far as NEC violation is concerned, the difference between discrete and continuous momentum can play an important role.

In Section \ref{semiclassical} we describe the semi-classical approximation in which the partition function is derived, to take into account the different saddle points which are relevant to tunnelling: 
static saddle points and the instanton/anti-instanton dilute gas. Details of the calculations with continuous momentum are given in 
Appendices \ref{AppStatic} and \ref{AppGas}. 
Section \ref{groundstate} focuses on the ground state of the effective action, with a non-extensive energy density providing the origin of NEC violation. 
The maximum effect occurs at zero temperature and is the regime in which we introduce corrections 
arising from discrete momentum in Section \ref{Casimir} via the Casimir energy. 
We find that tunnelling and the Casimir effect compete when the typical size of the box containing 
the field is of the order of the Compton wavelength of the corresponding particle. 
For a larger box, the Casimir effect seems to be dominant.

To summarise our results: in the low-temperature regime, the sum of energy density $\rho$ and pressure $p$ can be written in the form 
\be
\rho+p\simeq~A_\text{finite-T}~+~B_\text{tunnelling}~+~C_\text{Casimir}~,\nonumber
\ee
where
\begin{itemize}
\item the finite-temperature contribution $A$ is always positive (and vanishes exponentially for $T\to0$);
\item the tunnelling contribution $B$, calculated with continuous momentum, is always negative (and vanishes exponentially for $V\to\infty$);
\item the discrete momentum correction $C$ has a sign which depends on the geometry and topology of the finite box containing the field
(and vanishes for $V\to\infty$).
\end{itemize}
As expected, the NEC is satisfied at zero temperature and for infinite volume, 
where $\rho+p=0$ for a homogeneous vacuum.

\section{Semi-classical approximation}\label{semiclassical}

\subsection{Model}

Consider a single real scalar field $\phi(t,\mb{x})$ in Euclidean space, at finite-temperature $T=1/\beta$ and in a three-dimensional spacial volume $V$, 
described by the Euclidean action
\be \label{TAction}
\int_0^\beta dt\int_{V} d^3x\left(\frac{1}{2}(\partial\phi)^2+\frac{\lambda}{24}(\phi^2-v^2)^2+j\phi\right)~.
\ee
The finite volume is represented by a physical box containing the scalar field, in which 
we assume continuous momenta to calculate quantum corrections. Section \ref{Casimir}
discusses corrections arising from discrete momentum and the boundary conditions the field satisfies at the walls of the box.
Finite temperature requires field configurations to have periodic boundary conditions in Euclidean time and, as later 
discussed, has an impact on the saddle point configurations which are allowed in the partition function.

Introducing the dimensionless variables $\tau\equiv\omega t$ and
\be
\omega\equiv v\sqrt{\frac{\lambda}{6}}~~,~~\varphi\equiv\sqrt{\frac{\lambda}{6}}\frac{\phi}{\omega}~~,~~
k\equiv\sqrt{\frac{\lambda}{6}}\frac{j}{\omega^3}~,
\ee
leads to the bare action
\bea
S[\varphi]&=&\frac{\lambda v^4}{12\omega}\int_0^{\omega\beta} d\tau\int_{V} d^3x\Big((\varphi')^2+\frac{1}{\omega^2}(\nabla\varphi)^2\\
&&~~~~~~~~~~~~~~~~+\frac{1}{2}(\varphi^2-1)^2+2k\varphi\Big)~,\nonumber
\eea
where a prime represents a derivative with respect to the dimensionless Euclidean time $\tau$. 
As shown further in this article, the effective action is convex and we thus focus on the true vacuum, which occurs for vanishing source $j=0=k$.
As a consequence, no bubbles of true/false vacuum can form as they would have an infinite radius \cite{Coleman:1977py,Callan:1977pt}. 
We are therefore interested in time-dependent instantons only, 
beyond the static and homogeneous saddle points. The corresponding equation of motion is then 
\be \label{EoM}
\varphi^{\prime \prime}-\varphi^3+\varphi-k=0~,
\ee
where the solutions to this equation, $\varphi_i(\tau,\mb{x})$, are the saddle points of the partition function to be introduced below.

\subsection{Static saddle points}

Introducing the critical dimensionless source 
\be
k_c\equiv 2/(3\sqrt{3})~, 
\ee
allows us to distinguish two cases.

For $|k|>k_c$, there is only one static and homogeneous (real) solution to the equation of motion \eqref{EoM}, 
and quantisation of the theory can therefore be based on one saddle point only, leading to the usual 1PI effective potential. 

For $|k|<k_c$, the regime we focus on, there are two such solutions
\bea \label{Saddle points}
\varphi_L(k)&=&\frac{2}{\sqrt3}\cos\Big(\pi/3-(1/3)\arccos(k/k_c)\Big)\\
\varphi_R(k)&=&\frac{2}{\sqrt3}\cos\Big(\pi-(1/3)\arccos(k/k_c)\Big)\nn
&=&-\varphi_L(-k)~.\nonumber
\eea
The actions for these configurations are
\bea
S_L&\equiv & S[\varphi_L(k)]=B\omega\beta\left(4k-k^2+{\cal O}(k^3)\right)\\
S_R&\equiv & S[\varphi_R(k)]=S[\varphi_L(-k)]~,\nonumber
\eea
where
\be
B\equiv\frac{\lambda v^4 V}{24\omega}~.
\ee

\subsection{Instanton/anti-instanton gas}

In Euclidean time, and with the absence of a source, the motion described by equation (\ref{EoM}) corresponds to the motion 
in real-time with the upside-down potential $V(\varphi)\equiv-(\varphi^2-1)^2/2$, 
for which the minimum action $S_\text{inst}$ is obtained by the known solution
\be\label{instanton}
\varphi_\text{inst}(\tau)=\pm\tanh\left(\frac{\tau-\tau_0}{\sqrt{2}}\right)~,
\ee
where $0\le\tau_0\le\omega\beta$, and 
\be
S_\text{inst}\equiv S[\varphi_\text{inst}]=\frac{8\sqrt2}{3} B~.
\ee
Because of finite temperature though, field configurations should be periodic in Euclidean time, 
such that one needs to consider an instanton/anti-instanton pair as the basic building block.
For two distant ``jumps" at $\tau_1$ and $\tau_2$ such that $|\tau_1-\tau_2|\gg1$, the configuration can be approximated by \cite{Alexandre:2022qxc}
\be \label{Pair}
\varphi_\text{pair}(\tau)\simeq-\tanh\left(\frac{\tau-\tau_1}{\sqrt{2}}\right)\tanh\left(\frac{\tau-\tau_2}{\sqrt{2}}\right)~,
\ee
with an action exponentially close to $2S_\text{inst}$. 

In the presence of a source, the basic building block is in principle either a bounce or a shot (see \cite{Andreassen:2016cvx,Ai:2019fri} for reviews). 
However, since we are interested in the limit of vanishing source and periodic boundary conditions, 
the fundamental saddle point we consider behaves as the function (\ref{Pair}).
Assuming the jumps occur over a short time in comparison to $\beta$,
the instanton/anti-instanton pair spends the same time $\beta/2$ exponentially close to each 
static saddle point, resulting in an action for such a pair $\varphi_\text{pair}$ of
\be
S_\text{pair}\simeq \frac{1}{2}S_L+\frac{1}{2}S_R+2S_\text{inst}~.
\ee

\begin{figure}[h!]
     \centering
     \begin{subfigure}[b]{0.5\textwidth}
         \centering
         \includegraphics[width=\textwidth]{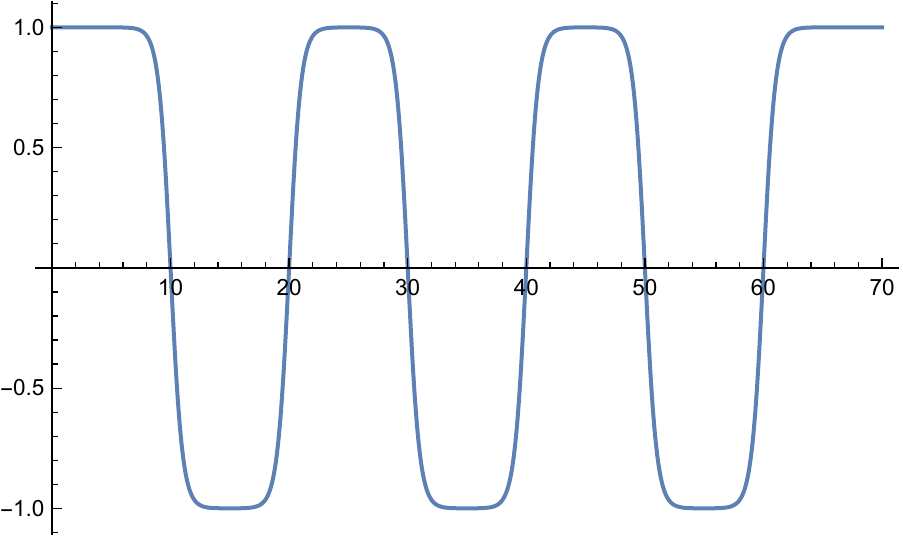}
         \caption{An exact saddle point configuration with 3 instanton/anti-instanton pairs and action $S_\text{3 pairs}$:
         the oscillations are periodic.} 
         \label{exactinst}
     \end{subfigure}
     \hfill
     \begin{subfigure}[b]{0.5\textwidth}
         \centering
         \includegraphics[width=0.945\textwidth]{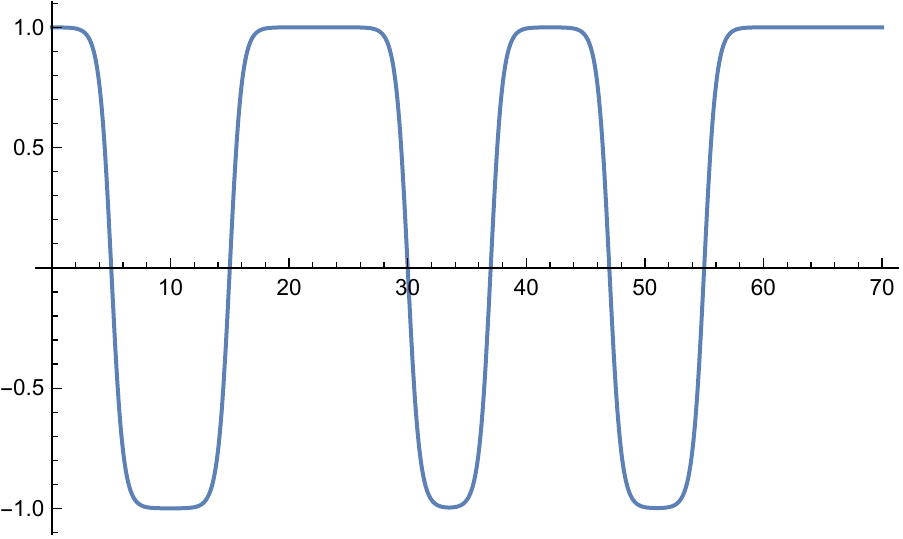}
         \caption{An approximate saddle point configuration with 3 instanton/anti-instanton pairs:
         the jumps are randomly distributed, but the average distance between them is larger than their width, 
         such that they keep their shape and the action of the configuration is also $S_\text{3 pairs}$.}
         \label{approxinst}
     \end{subfigure}
        \caption{Examples of exact and approximate saddle points. In the dilute gas approximation, the difference between the corresponding actions 
        is of order $B\omega\beta\exp(-\omega\beta)\ll1$, and the partition function is dominated by the whole set of approximate saddle points.}
\end{figure}

Revisiting the analogy of classical mechanics in the upside-down potential $V(\varphi)$, 
the other possible saddle points consist of periodic oscillations made of $n$ instanton/anti-instanton pairs, 
where the value of $n$ depends on how ``exponentially close" the oscillations from a static saddle point begins. An example of an exact saddle point is given 
in Fig.\ref{exactinst}.
Assuming the total Euclidean time $\beta$ is large enough
to leave the structure of pairs intact, the time spent close to one static saddle point is the same as the time spent close to the other and the total 
action for $n$ pairs is
\be\label{Snpairs}
S_\text{n pairs}\simeq \frac{1}{2}S_L+\frac{1}{2}S_R+2nS_\text{inst}~.
\ee

The latter ``crystalline" structure, with $n$ periodic oscillations, corresponds to an exact solution of the equation of motion. 
For large $\beta$, where the average distance between instantons and anti-instantons remains 
large compared to their width, a translation of each jump leaves the action $S_\text{inst}$ invariant
and the resulting highly degenerate ``gas" of instanton/anti-instanton pairs dominates the partition function. 
An example of an approximate saddle point is given in Fig.\ref{approxinst}. In this ``dilute gas" approximation, the 
$n$ instanton/anti-instanton pair configurations spend on average an equal time $\beta/2$ close to each static saddle point, with the same total action (\ref{Snpairs}) as 
for an exact $n$-pair configuration as a result of the translational invariance of jumps.

\subsection{Partition function}

The partition function is evaluated in the semi-classical approximation via a sum over the two static saddle points, $\varphi_L$ and $\varphi_R$, 
and the dilute gas of $n$ instanton/anti-instanton pairs for all possible values of $n$.
Together, with the corresponding one-loop fluctuation factors $F_{L,R}$ and $F_n$,
the semi-classical approximation of the partition function reads
\bea\label{ZTotal}
Z[k]&\simeq&F_L(\beta)\exp(-S_L)+F_R(\beta)\exp(-S_R)\\
&+&\sum_{n=1}^\infty \left(\prod_{i=1}^{2n}\int_{\tau_{i-1}}^{\omega_R\beta}\mathrm{d}\tau_i\right) F_n\exp(-S_\text{n pairs})~.\nonumber
\eea
In the latter expression, the product of integrals over the times $\tau_i$ where the jumps occur 
corresponds to the zero-mode of the fluctuation factor for the saddle point made of $n$ instanton/anti-instanton pairs.
Indeed, the translational invariance of the action means that the $i_\text{th}$ jump can happen at any time $\tau_i\in[\tau_{i-1},\beta]$.
For finite temperature though, there is a maximum number of instanton/anti-instanton pairs, however, the error made in the summation for $n\to\infty$
is negligible since each term is suppressed by $\exp(-nS_\text{inst})$.
With the fluctuation factors derived in Appendix \ref{AppStatic} and Appendix \ref{AppGas}, we can write 
\bea\label{ZSigma}
Z[k]&=&\exp\Big(-\Sigma_L(\beta)\Big)+\exp\Big(-\Sigma_R(\beta)\Big)\nn
&&+\exp\Big(-\Sigma_\text{gas}(\beta)\Big)~,
\eea
where $\Sigma_L,\Sigma_R$ and $\Sigma_\text{gas}$ are the connected graphs generating functionals for the static saddle points 
and the gas of instanton/anti-instanton pairs respectively.
We note that an instanton or anti-instanton does not lead to any imaginary part in the partition function,
unlike a bounce, since the former are monotonous functions of the Euclidean time, such that the fluctuation operator does not have negative eigenvalues \cite{Kleinert:2004ev}.

\subsubsection{Static saddle points}

One-loop quantum corrections can be split into two contributions: the zero-temperature corrections, containing all the divergences, and the divergence free finite-temperature dependent corrections.
The zero-temperature contribution is calculated in \cite{Alexandre:2022sho} and is expressed in terms of the renormalised parameters. 
It is mentioned here that, in the case of several saddle points and in order to avoid confusion between loop orders, 
renormalisation should be done at the level of the individual connected graphs generating functionals 
before performing the Legendre transform. The finite-temperature contribution can be calculated using the Schwinger proper time representation
- see Appendix \ref{AppStatic} - and the overall contribution is
\bea\label{SigmaLR}
&&\Sigma_{L,R}(\beta)\\
&=& B_r\omega_r\beta\Bigg((\varphi_{L,R}^2-1)^2+4k\varphi_{L,R}\nn
&&+\frac{\lambda_r}{96\pi^2}(3\varphi_{L,R}^2-1)^2\ln\left(\frac{3}{2}\varphi_{L,R}^2-\frac{1}{2}\right)\nn
&&-\frac{\lambda_r(3\varphi_{L,R}^2-1)}{3\pi^2}\sum_{l=1}^\infty \frac{K_2\left(l\omega_r\beta\sqrt{3\varphi_{L,R}^2-1}\right)}{(l\omega_r\beta)^2}\Bigg)~.\nonumber
\eea
In the previous expression, the renormalised parameters are
\bea
\lambda_r&\equiv&\lambda-\frac{3  \lambda^2}{32 \pi^2} \log \left(\frac{\Lambda^2}{\lambda v^2}\right)~,\\
v_r^2&\equiv&v^2-\frac{3 \Lambda^2}{16 \pi^2}+\frac{ \lambda v^2}{16 \pi^2} \log \left(\frac{\Lambda^2}{\lambda v^2}\right)~,\nn
B_r&\equiv&\frac{\lambda_r v_r^4 V}{24\omega_r}~,\nn
\omega_r&\equiv&v_r\sqrt{\frac{\lambda_r}{6}}~,\nonumber
\eea
and $K_2(z)$ is a modified Bessel function of the second kind with asymptotic behaviour
\be\label{asymptotiK}
K_2(z\to\infty)\simeq e^{-z}\sqrt{\frac{\pi}{2z}}~.
\ee
We note that $l$ does not correspond to Matsubara modes.
Also, the temperature-independent part of the expression (\ref{SigmaLR}) reproduces the 
zero-temperature result derived in \cite{Alexandre:2022sho}.

\subsubsection{Gas of instanton/anti-instanton pairs}

The evaluation of $\Sigma_{gas}$ involves the fluctuation factor above each jump 
and includes a summation over the allowed jump positions in the interval $\tau\in[0,\beta]$ \cite{Kleinert:2004ev}. 
The additional contribution of quantum fluctuations arises from the ``flat" parts of the instanton/anti-instanton configurations, 
which are exponentially close to each static saddle point for the approximate average time of $\beta/2$ when neglecting the width of each jump compared to $\beta$. Performing the resummation over instantons/anti-instantons,
we show in Appendix \ref{AppGas} that the corresponding connected graphs generating functional is then
\bea\label{Sigmagas}
\Sigma_\text{gas}(\beta)&\simeq&\Sigma_L(\beta/2)+\Sigma_R(\beta/2)\\
&&~-\ln\Big(\cosh(\bar{N})-1\Big)~,\nonumber
\eea
where
\be\label{Nbar}
\bar{N}\equiv \omega_r\beta \sqrt{\frac{6}{\pi}S_\text{inst}}~e^{-S_\text{inst}}~,
\ee
corresponding to the average number of instanton/anti-instanton pairs at temperature $T=1/\beta$.
In this article we are interested in the limit $\omega_r\beta\gg1$ for a fixed volume - and thus fixed action $S_\text{inst}$ - 
such that we consider the situation where $\bar N\gg1$, corresponding to the full tunnelling regime. 
In the situation where $\beta$ is fixed and $V$ becomes large we have $\bar N\ll1$, 
where tunnelling is suppressed and the system is better approximated by SSB \cite{Alexandre:2022qxc}.

\section{Non-extensive ground sate}\label{groundstate}

\subsection{One-particle-irreducible effective action}

From the partition function evaluated for a constant source $j$, the classical field is obtained as
\be
\phi_c\equiv-\frac{1}{Z}\frac{\delta Z}{\delta j}~\to~-\frac{1}{V\beta Z}\frac{\partial Z}{\partial j}~,
\ee
which, in terms of the dimensionless quantities previously introduced, can be written as
\be
\varphi_c=-\frac{1}{4B_r\omega_r\beta Z} \frac{\partial Z[k]}{\partial k}~.
\ee
From the expression (\ref{ZSigma}) for the partition function, together with the expressions (\ref{SigmaLR}) and (\ref{Sigmagas}),
the classical field is expanded in powers of the source $k$
\be
\varphi_c=\left(-f_0+\frac{\lambda_r}{128\pi^2}f_1\right)k +\Od(k^3)~,
\ee
where
\bea
f_0&\equiv& \frac{1+16B_r\omega_r\beta+\cosh(\bar{N})}{2\Big(1+\cosh(\bar{N})\Big)}\\
f_1&\equiv& \frac{7+32B_r\omega_r\beta+7\cosh(\bar{N})}{1+\cosh(\bar{N})}~.\nonumber
\eea
Consistently with the symmetry of the bare potential, the classical field $\phi_c$ is an odd function of $k$: 
the even powers of $k$ cancel out in the expression for $\varphi_c$ after adding the contribution of the different saddle points,
leading to the mapping $k=0\Leftrightarrow\varphi_c=0$.

We then perform the Legendre transform, after expressing the source as a function of the classical field
\be 
k(\varphi_c)=-\frac{1}{2}\left(g_0+\frac{\lambda}{16\pi^2} g_1\right)\varphi_c +\Od(\varphi_c^3)~,
\ee
where 
\bea
g_0&\equiv& \frac{4\Big(1+\cosh(\bar{N})\Big)}{1+16B_r\omega_r\beta+\cosh(\bar{N})}\\
g_1&\equiv& \frac{\Big(1+\cosh(\bar{N})\Big)\Big(7+32B_r\omega_r\beta+7\cosh(\bar{N})\Big)}{\Big(1+16B_r\omega_r\beta+\cosh(\bar{N})\Big)^2}~.\nonumber
\eea
The effective action for a constant configuration is finally
\bea\label{Gamma}
\Gamma(\varphi_c)&=&-\ln Z\Big(k(\varphi_c)\Big)-4 B_r\omega_r\beta\int k(\varphi_c)~d\varphi_c\\
&=&\Gamma(0)+ B_r\omega_r\beta\left(g_0+\frac{\lambda}{16\pi^2} g_1\right)\varphi_c^2+\Od(\varphi_c^4)~,\nonumber
\eea
where
\bea\label{Gamma0}
\Gamma(0)&=&-\ln Z(0)\\
&=&-\ln\left(2e^{-\Sigma_0(\beta)}+ e^{-2\Sigma_0(\beta/2)}\Big(\cosh(\bar{N})-1\Big)\right)~,\nonumber
\eea
and $\Sigma_0\equiv\Sigma_L|_{k=0}=\Sigma_R|_{k=0}$. The effective potential $U_\text{eff}$ is finally given by 
\be
\Gamma(\phi_c)=V\beta U_\text{eff}(\phi_c)~,
\ee
and, as expected, it satisfies the following properties: 
\begin{itemize}
    \item it is a convex function of $\phi_c$, since the mass term is positive;
    \item the ground state is at $\varphi_c=0$, or equivalently $k=0$; 
    \item it has a non-trivial volume-dependence and is thus non-extensive.
\end{itemize} 
For the following studies of NEC violation we focus on the ground state $\varphi_c=0$.

\subsection{NEC violation}

\begin{figure}[h!]
     \centering
     \includegraphics[width=0.45\textwidth]{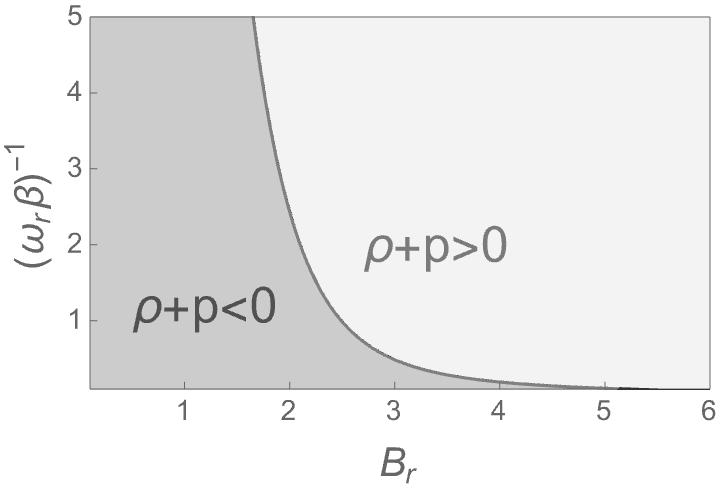}
    \caption{The boundary between the regions where the NEC is satisfied and where it is violated 
    due to the competition of tunnelling and thermal fluctuations. The plot shows the curve $V(T)$ in terms
    of the dimensionless variables used in this article.} 
    \label{rho+p=0}
\end{figure}

The ground state density $\rho$ and pressure $p$ are obtained from the free energy
\be
F=\frac{1}{\beta}\Gamma(0)=-\frac{1}{\beta}\ln Z(0)~,
\ee
and their sum can be written as \cite{Alexandre:2022qxc}
\bea
\rho+p &=&\frac{1}{V}\left(F-T \frac{\partial F}{\partial T}\right)-\frac{\partial F}{\partial V} \nn
&=&-T \frac{\partial U_\text{eff}(0)}{\partial T}-V \frac{\partial U_\text{eff}(0)}{\partial V}~.
\eea
From the expression (\ref{Gamma0}), we obtain for $\omega_r\beta\gg1$
\bea\label{rho+p}
\rho+p&\simeq& \frac{4\omega_R^{5/2}}{(\sqrt2\pi\beta)^{3/2}}~e^{-\omega_R\beta/\sqrt{2}}\\
&&-\frac{\omega_R}{V}\left(S_\text{inst}+\frac{1}{2}\right)\sqrt{\frac{6}{\pi}S_\text{inst}}~e^{-S_\text{inst}}~.\nonumber
\eea
On the right-hand side, the first term corresponds to thermal fluctuations and the second term corresponds to tunnelling. 
These terms compete for the overall sign of $\rho+p$ leading to the following cases:

\begin{itemize} 

\item {\it Infinite volume: $\rho+p\ge0$}\\
In the limit of infinite volume tunnelling is suppressed, as seen via the vanishing of the average number (\ref{Nbar}) of instanton/anti-instanton pairs
for any fixed temperature: $\lim_{V\to\infty}\bar{N}=0$ for fixed $\beta$. Hence only thermal fluctuations contribute and 
\be
\rho+p=\frac{4\omega_R^{5/2}}{(\sqrt2\pi\beta)^{3/2}}~e^{-\omega_R\beta/\sqrt{2}}~,
\ee
with $\rho+p\to0$ as the temperature goes to 0 or equivalently $\beta\to\infty$. 
This result is not surprising: the limit of infinite volume corresponds to SSB and, as expected, the NEC is satisfied;

\item{\it Finite volume and zero temperature: $\rho+p<0$}\\
In this situation, only the tunnelling term contributes and 
\be
\rho+p=-\frac{\omega_R}{V}\left(S_\text{inst}+\frac{1}{2}\right)\sqrt{\frac{6}{\pi}S_\text{inst}}~e^{-S_\text{inst}}~.
\ee 
The NEC is violated as a consequence of the explicit volume-dependence of the effective potential;

\item{\it Boundary $\rho+p=0$}\\
We sketch in Fig.\ref{rho+p=0} the boundary $V(T)$ 
between the region where the NEC is satisfied and the region where the NEC is violated.

\end{itemize}

Finally, we note that NEC violation is suppressed exponentially with the volume, 
unlike the power law suppression which is found with $O(4)$-symmetric Euclidean 
spacetime coordinates \cite{Alexandre:2022sho,Alexandre:2019ygz,Alexandre:2021imu}.

\section{Discrete momentum corrections}\label{Casimir}

We focus here on the ground state obtained for $k=0$ in the case of zero temperature, where NEC violation arising from tunnelling is maximum. 

The previous sections ignore quantisation of momentum when calculating the connected graphs generating functional for each static saddle point in a finite volume.
As we explain below, the evaluation of $\Sigma_{L,R}$ with discrete momentum consists of taking into account the relevant Casimir energy.
There is no such contribution from the jumps in the instantons/anti-instantons since the corresponding one-loop corrections do not depend on momentum.

\subsection{Vacuum energy}

The Casimir contribution to the connected graphs generating functional is defined as
\be
\Sigma_\text{Cas}\equiv\Sigma_{L,R}|_\text{discrete}-\Sigma_{L,R}|_\text{continuum}~,
\ee
where the ultraviolet divergences cancel out since they are identical in the discrete and continuum cases. 
For zero temperature and vanishing source, the expression (\ref{SigmaLR}) gives 
\be
\Sigma_{L,R}(k=0,T=0)|_\text{continuum}=\lim_{\beta\to\infty}\Sigma_0(\beta)=0~,
\ee
such that, instead of eq.(\ref{Gamma0}), one-loop corrections obtained with discrete momentum lead to 
\bea
\Gamma(0)
&=&-\ln\left(2e^{-\Sigma_\text{Cas}}+ e^{-\Sigma_\text{Cas}}\Big(\cosh(\bar{N})-1\Big)\right)\nn
&=&~\Sigma_\text{Cas}-\ln\Big(\cosh(\bar{N})+1\Big)~.
\eea
The above expression takes advantage of the proportinality between $\Sigma_0$ and $\beta$ in the limit of vanishing temperature, such that
\be
2\Sigma_0(\beta/2)\to\Sigma_0(\beta),
\ee
as $\beta\to\infty$. In the situation of one saddle point, and therefore no tunnelling, $\Gamma(0)=\Sigma_\text{Cas}=\beta E_\text{Cas}$ 
where $E_\text{Cas}$ is the Casimir energy corresponding to quantum fluctuations about a single vacua $\pm v$ 
(where one has approximately quadratic fluctuations with mass $m=\sqrt2\omega_r$). Hence 
\be
U_\text{eff}(0)=\frac{E_\text{Cas}}{V}-\frac{1}{V\beta}\ln\Big(\cosh(\bar{N})+1\Big)~,
\ee
and we see the additive nature of the Casimir effect and tunnelling contributions, 
similarly to the finite-temperature contribution. The sum of density and pressure reads finally
\bea
\rho+p&=&\frac{E_\text{Cas}}{V}-\frac{\partial E_\text{Cas}}{\partial V}\\
&&-\frac{\omega_R}{V}\left(S_\text{inst}+\frac{1}{2}\right)\sqrt{\frac{6S_\text{inst}}{\pi}}~e^{-S_\text{inst}}~.\nonumber
\eea

\subsection{Casimir contribution to the NEC}

The Casimir energy is highly sensitive to the geometry of the box containing the field, as well as 
the boundary conditions used on the corresponding surfaces \cite{Bordag:2001qi}. 
For a scalar field $\varphi(t,x)$ in the interval $x\in[0,L]$ for example, the possible choices of boundary conditions are defined as follows
\bea
\text{Dirichlet:}~~& &\varphi(t,0)=\varphi(t,L)=0\\
\text{Neumann:}~~& &\partial_x\varphi(t,0)=\partial_x\varphi(t,L)=0\nn
\text{Periodic:}~~& &\varphi(t,0)=\varphi(t,L)~.\nonumber
\eea
For the cases we consider, the asymptotic form of the Casimir effect is identical for both Dirichlet and Neumann boundary conditions. We thus consider mixed boundary conditions, where different subsets of the boundary can possess either Dirichlet or Neumann conditions. 
For the case of mixed boundary conditions, the Casimir energy is dependent on the size/curvature of the material boundaries, and for the case of periodic boundary conditions, it is dependent on the period length/curvature of the non-trivial spacetime. 
A `general rule' states that flat geometries lead to exponential suppression of the Casimir energy for $mL\gg1$, 
where $L$ is the length scale of the relevant boundaries, and that curved geometries lead to power-law 
suppression of the Casimir energy for $mR\gg1$, where $R$ is the radius of curvature of the relevant surfaces.
There are exceptions to this general rule though, which are highlighted in the following examples.\\

$\bullet$ {\it Dirichlet boundary conditions, flat boundaries}\\
The original Casimir configuration consists of a scalar field constrained between two parallel, flat mirrors with surface area $A$ and separation $a$, with the scalar field satisfying Dirichlet conditions on the boundaries. The corresponding Casimir energy is \cite{Bordag:2001qi}
\be
E_\text{Cas}\simeq\left\{
\begin{array}{c}
-\frac{A\pi^2}{1440a^3} ~~~~~~~~~~~~~~~~~~~~\mbox{for}~~~~am\ll1\\
-\frac{A}{8\sqrt2}\left(\frac{m}{\pi a}\right)^{3/2}~e^{-2ma}~~~~\mbox{for}~~~~am\gg1
\end{array}\right.
\ee
and is always negative.\\

$\bullet$ {\it Dirichlet boundary conditions, curved boundaries}\\
For dimensional reasons, the Casimir energy for a scalar field confined within the curved boundary of a 2-sphere of radius $R$ with Dirichlet boundary conditions is given in terms of the dimensionless function
\be 
E_\text{Cas}=\frac{1}{R}f(mR)~,
\ee
and is found to obey power law suppression in $mR$, for $mR\gg1$ \cite{Kirsten:1999qjn}.\\

$\bullet$ {\it Periodic boundary conditions, flat spacetime}\\
For a scalar field confined to the surface of a 3-torus (a rectangular box with periodic boundary conditions), the sign of the Casimir energy depends 
on the ratio of the lengths of the box and we have \cite{Lim:2008yv}
\be
E_\text{Cas}\simeq-\frac{(mL)^{3/2}}{L}\exp(-mL)~~~~\mbox{for}~~~~mL\gg1~,
\ee
where $L$ is the typical size of the period length.\\

$\bullet$ {\it Periodic boundary conditions, curved spacetime}\\
For a scalar field confined to the surface of a 3-sphere with radius $R$, we would expect the asymptotic form to be a power law in R. However, this special case is an exception to the general rule as a consequence of the accidental vanishing of the heat-kernel coefficients (see Sec. 3 of \cite{Bordag:2001qi} for details). The resulting Casimir energy has instead an exponential asymptotic form, as in the case of flat geometries \cite{Mamaev:1976zb}
\be
E_\text{Cas}\simeq+\frac{(mR)^{5/2}}{R}\exp(-2\pi mR)~~~~\mbox{for}~~~~mR\gg1~.
\ee
\\

\begin{figure}[h!]
     \centering
     \includegraphics[width=0.45\textwidth]{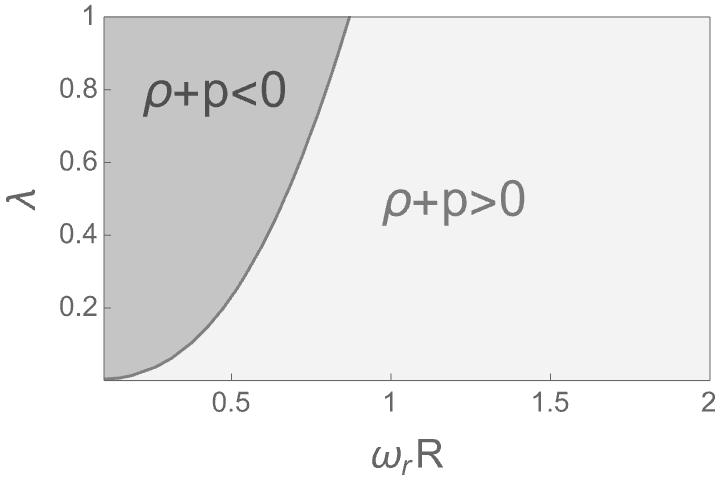}
    \caption{The boundary between the regions where the NEC is satisfied and where it is violated, due to the competition of the Casimir energy and tunnelling 
    at zero temperature on a 3-sphere. 
    The plot shows the curve $R(\lambda)$ in terms of the dimensionless variables used in this article.} 
    \label{fig: rho+p=0 Casimir}
\end{figure}

The above examples display how the Casimir effect for a massive scalar field is at most suppressed by 
the exponential $e^{-mL}$, where $L$ is a typical size of the boundary containing the field.
On the other hand, the tunnelling contribution to the NEC, calculated with continuous momentum, is proportional to
\be
e^{-S_\text{inst}}\sim\exp\left(-\frac{(mL)^3}{\lambda}\right)~, 
\ee
and is therefore negligible compared to the Casimir contribution in the regime 
$mL\gg\sqrt\lambda$. For $mL\sim\sqrt\lambda$ though, 
tunnelling competes with the Casimir effect and can change the sign of $\rho+p$ in the situation where the Casimir
energy is positive. As an example, we sketch in Fig.\ref{fig: rho+p=0 Casimir} the boundary $R(\lambda)$ 
between the region where the NEC is satisfied and the region where it is violated, 
due to the competition between tunnelling and the Casimir effect on a 3-sphere.

We note however two important points regarding the Casimir examples cited here:
{\it(i)} they are valid for ideal surfaces only, 
and a realistic confining mechanism for the scalar field would lead to a modification of the Casimir vacuum energies,
especially if the field is confined by an external potential instead of a physical box \cite{Graham:2002yr};
{\it(ii)} they assume free scalar fields and ignore its self-interactions.
On the other hand, the tunnelling mechanism described here: {\it(i)} necessitates the field to be self-interacting; 
{\it(ii)} is not sensitive to the geometry/topology of the box containing the field. Hence the conclusions regarding 
which effect dominates could be modified by a more thorough study, depending on the situation which is considered.

Finally, the Average Null Energy Condition is not violated by the present mechanism. 
Indeed, if we take into account the energy necessary to maintain the confining mechanism the overall ground state 
of the system does not violate the NEC \cite{Sopova:2002cs}, consistently with what is expected from causality \cite{Hartman:2016lgu}.

\section{Conclusions}

Tunnelling between degenerate vacua is exponentially suppressed with the volume of the box containing the field, 
but nevertheless allows the possibility of NEC violation at low temperatures. Taking into account discrete momentum of fluctuations in a finite volume 
implies this effect is mainly relevant for situations where the typical size of the box is not too large compared to the Compton wave length of the particle, and where tunnelling can lead to an overall NEC violation. A potential application lies in axion physics, 
where the de Broglie wavelength can be of order 1 kpc \cite{Marsh:2015xka}
with the confinement provided by a gravitational well.

Exponential suppression in the volume could potentially be avoided by a consideration of non-degenerate vacua, 
where other saddle points with a volume-independent action become relevant, as in the original study of false vacuum decay \cite{Coleman:1977py,Callan:1977pt}. 
The resulting effective action would be non-extensive in a certain regime of the classical field, 
but more studies need to be done for the status of NEC violation in the corresponding vacuum.

Finally, NEC violation could play an important role in Early Universe Cosmology, where tunnelling could provide a 
dynamical mechanism for a cosmological bounce, as explained in \cite{Alexandre:2019ygz,Alexandre:2021imu}: 
as the Universe contracts, tunnelling switches on and violates the NEC, which induces a bounce after which tunnelling is suppressed as the Universe expands.
This scenario necessitates the study of tunnelling in a Friedman-Lemaitre-Robertson-Walker background though, and is left for future work.

\section*{Acknowledgements}
The authors would like to thank Klaus Kirsten for valuable correspondence regarding the Casimir effect,  
and JA would like to thank Janos Polonyi for enlightening discussions.
This work is supported by the Leverhulme Trust (grant RPG-2021-299) and the Science and Technology
Facilities Council (grant STFC-ST/T000759/1). For the purpose of Open Access, 
the authors have applied a CC BY public copyright licence to any Author Accepted Manuscript version arising from this submission.

\appendix

\section{Fluctuation factor for a static saddle point}\label{AppStatic}

The fluctuation factors for the static saddle points are calculated with continuous 3-dimensional momenta, introducing the cut-off $\Lambda$ in the 
Schwinger proper time representation of the propagator. Introducing the dimensionless Matsubara frequency $\nu_n\equiv2\pi n/\omega\beta$, we have
\bea
&&\mbox{Tr}\left\{\ln\left(\delta^2S[\varphi_i]\right)\right\}\\
&=&V\int\frac{d^3p}{(2\pi)^3}\sum_{n=-\infty}^\infty\int_{1/\Lambda^2}^{\infty}
\frac{\mathrm{d}s}{s}\mathrm{e}^{-4B\omega\beta s(p^2/\omega^2+\nu_n^2+3\varphi_i-1)}\nn
&=& \frac{V\omega^3}{2\pi^2}\sum_{n=-\infty}^\infty\int_{1/X^2}^{\infty}\frac{\mathrm{d}x}{x}\int_0^\infty\mathrm{d}q~q^2\mathrm{e}^{-x(q^2+\nu_n^2+3\varphi_i-1)}\nn
&=& \frac{V\omega^3}{8\pi^{3/2}}\sum_{n=-\infty}^\infty\int_{1/X^2}^{\infty}\frac{\mathrm{d}x}{x^{5/2}}\mathrm{e}^{-x(\nu_n^2+3\varphi_i-1)}\nn
&=& \frac{V\omega^3}{8\pi^{3/2}}\int_{1/X^2}^{\infty}\frac{\mathrm{d}x}{x^{5/2}}\mathrm{e}^{-x(3\varphi_i-1)}\vartheta_0\left(\frac{4\pi x}{\omega^2\beta^2}\right)~,\nonumber
\eea
where the dimensionless variables are
\be
q\equiv \frac{p}{\omega}\quad,\quad
x\equiv 4B\omega\beta s\quad,\quad
X^2\equiv\frac{\Lambda^2}{4B\omega\beta}~,
\ee
and $\vartheta_0(y)$ is the Jacobi function
\be 
\vartheta_0(y)\equiv\sum_{n=-\infty}^\infty\mathrm{e}^{-\pi yn^2}~.
\ee 
Making use of the following property
\be 
\vartheta_0(y)=y^{-1 / 2} \vartheta_0(1 / y)~,
\ee
the above becomes
\bea
&&\mbox{Tr}\left\{\ln\left(\delta^2S[\varphi_i]\right)\right\}\\
&=& \frac{V\omega^4\beta}{16\pi^2}\int_{1/X^2}^{\infty}\frac{\mathrm{d}x}{x^3}\mathrm{e}^{-x(3\varphi_i-1)}\vartheta_0\left(\frac{\omega\beta}{4\pi x}\right)\nn
&=& \frac{V\omega^4\beta}{16\pi^2}\int_{1/X^2}^{\infty}\frac{\mathrm{d}x}{x^3}\mathrm{e}^{-x(3\varphi_i-1)}\sum_{n=-\infty}^\infty e^{-\omega^2\beta^2n^2/4x}\nn
&=& \lambda\frac{B\omega\beta}{24\pi^2}\Big(I_\Lambda(\varphi_i)+I_T(\varphi_i)\Big)~,\nonumber
\eea
where
\bea 
I_\Lambda(\varphi_i)&\equiv& \int_{1/\Lambda^2}^{\infty}\frac{\mathrm{d}x}{x^3}~\mathrm{e}^{-x(3\varphi_i-1)}\\
I_T(\varphi_i) &\equiv& 2\sum_{n=1}^{\infty}\int_0^{\infty}\frac{\mathrm{d}x}{x^3}~\mathrm{e}^{-x(3\varphi_i-1)-\omega^2\beta^2n^2/4x}~.\nonumber
\eea
The first integral $I_\Lambda$ is the temperature-independent divergent integral which, after renormalisation, 
produces the same results as in the zero-temperature case \cite{Alexandre:2022sho}. 
The second integral $I_T$ is the temperature-dependent contribution corresponding to the finite-temperature corrections. 
It is finite, which is why the cut-off is taken to infinity in this specific term.
This temperature-dependent integral can be written in terms of the modified Bessel functions of the second kind $K_2(z)$ as 
\be 
I_T(\phi_i)= \sum_{n=1}^{\infty}\frac{16(3\varphi_i-1)}{(n\omega\beta)^2}K_2(n\omega\beta\sqrt{3\varphi_i-1})~.
\ee
Together with the integral $I_\Lambda$, the connected graphs generating functional for homogeneous saddle points is given by
eq.(\ref{SigmaLR}).

\section{Fluctuation factor for the instantons/anti-instantons gas}\label{AppGas}

We calculate here the contribution $\exp(-\Sigma_\text{gas})$ to the partition function (\ref{ZSigma}), 
following the known approach in studies of tunnelling effects \cite{Kleinert:2004ev}.

The invariance of the action for $n$ instanton/anti-instanton pairs under the translation of the jumps leads to the degeneracy factor 
in the partition function
\bea
\left(\prod_{i=1}^{2n}\int_{\tau_{i-1}}^{\omega\beta}\mathrm{d}\tau_i\right)=\frac{(\omega\beta)^{2n}}{(2n)!}~,
\eea
where $\tau_i\in[\tau_{i-1},\omega\beta]$ and $\tau_0=0$, since successive instanton jumps can only occur after previous ones.
Each jump has an associated fluctuation factor $\sqrt{6S_\text{inst}/\pi}$ and thus the total fluctuation factor is given by the 
product of the contributions of the ``flat" parts of the $n$-pairs of instanton/anti-instantons and the $n$ pairs of jumps.
On average, each configuration of $n$ instanton/anti-instanton pairs spends the same time $\simeq\beta/2$ close to each static saddle point,
such that the expression for $F_n$ is finally
\be
F_n=F_L(\beta/2)F_R(\beta/2)\left(\frac{6S_{int}}{\pi}\right)^n~.
\ee
Substituting the above results into the partition function \eqref{ZTotal}, along with the total action \eqref{Snpairs} for $n$ pairs, 
yields the total contribution to the partition function due to instanton/anti-instanton pairs
\bea
&&\exp(-\Sigma_\text{gas})\\
&=&e^{-\Sigma_L[\beta/2]}e^{-\Sigma_R[\beta/2]}\sum_{n=1}^\infty \frac{(\omega\beta)^{2n}}{(2n)!}\left(\frac{6S_\text{int}}{\pi}\right)^n e^{-2nS_\text{int}}\nn
&=&\exp\Big(-\Sigma_L[\beta/2]-\Sigma_R[\beta/2]\Big)\nn
&&\times\left(\cosh\left(\omega\beta\sqrt\frac{6S_\text{int}}{\pi}~e^{-S_\text{inst}}\right)-1\right)~,\nonumber
\eea
This leads to the expression (\ref{Sigmagas}), where the parameters can be replaced by their renormalised version, since the overall expression
is already at one-loop.

\bibliography{bibliography}

\end{document}